\documentclass{aastex631}
\usepackage{comment}
\usepackage{xcolor,colortbl}
\usepackage{threeparttable}

\begin{document}

\title{The Reawakening of 4U 1755$-$338 after 25 Years of Quiescence: Spectro-temporal Analysis Using Multi-instrument X-ray Data}

\email{geethuprabhakar3@gmail.com, geethuprabhakar.17@res.iist.ac.in}
\email{samir@iist.ac.in}

\author[0009-0001-0469-122X]{Geethu Prabhakar}
\affiliation{Department of Earth and Space Sciences \\
Indian Institute of Space Science and Technology (IIST), Trivandrum - 695547, India}

\author[0000-0001-9371-7104]{Samir Mandal}
\affiliation{Department of Earth and Space Sciences \\
Indian Institute of Space Science and Technology (IIST), Trivandrum - 695547, India}

\begin{abstract}
The black hole X-ray binary 4U 1755$-$338 underwent an outburst in 2020 after 25 years of quiescence.
The comprehensive spectral analysis revealed that the system has a low interstellar neutral hydrogen column density of $0.34\pm0.01 \times$10$^{22}$ cm$^{-2}$. The outburst began with a low mass-accretion rate and was characterized as a low-luminosity outburst. The radius of the inner accretion disc remained constant throughout the outburst. Additionally, a growing neutral medium with constant density was detected in the local environment of 4U 1755$-$338. The hardness–intensity diagram (HID) did not follow the standard q-shaped pattern, indicating a non-canonical outburst. Instead, the HID showed a correlated evolution of hardness and source flux, suggesting a thermal disc origin of the flux. A wideband spectral analysis was performed using simultaneous \textit{NICER-NuSTAR} data in two frameworks, based on \textit{kerrbb} and \textit{bhspec}. The results of \textit{bhspec (kerrbb)} based modeling indicate that 4U 1755$-$338 is a high-inclination system, $67.44_{-3.03}^{+9.75}$ ($75.25_{-4.68}^{+5.59}$) degrees, and harbors a moderately spinning black hole with a spin parameter of $0.78_{-0.14}^{+0.02}$ ($0.50_{-0.43}^{+0.19}$) and a mass of $3.37_{-1.04}^{+0.45}\ (3.28_{-1.1}^{+1.7})M_{\odot}$ respectively. The inferred key parameters: black hole mass, spin, and system inclination are consistent across both modeling approaches. No reflection features were detected in the spectra of 4U 1755$-$338. The high spectral index, the blackbody nature ($L\propto T^4$) of the hardness ratio, the absence of reflection signatures, and the weak variability in the power density spectra indicate that the source remained in the high/soft state throughout the outburst.
  
\end{abstract}


\section{Introduction} \label{sec:intro}
There have been decades of observations and studies aimed at understanding the physics of accretion in X-ray binaries (XRBs). Black hole X-ray binaries (BH-XRBs) consist of a stellar-mass black hole (BH) accreting matter from a companion star, emitting enormous amounts of X-rays in the process. BH-XRBs are known for their episodic outbursts if they are a low-mass X-ray binary (LMXB). During an outburst, the X-ray emission undergoes dramatic changes. These outbursts are triggered by variations in the accretion rate onto the black hole resulting from instabilities \citep{1974PASJ...26..429O,1996ApJ...464L.139V} in the accretion disc. Studying these outbursts allows astronomers to probe the dynamics of accretion flows and the mechanisms driving variability in X-ray emission.

The observed composite X-ray spectrum of a BH-XRB consists of several components, such as soft and hard X-ray components, reflection features, etc. The primary source of soft X-ray emission typically originates from blackbody radiation emitted by a geometrically thin and optically thick accretion disc \citep{1973A&A....24..337S}. This thermal emission undergoes inverse Compton scattering from a hot electron cloud (\textit{corona}) situated near the black hole \citep{1976ApJ...204..187S,1980A&A....86..121S,1995ApJ...455..623C}. The resulting Compton spectrum exhibits a powerlaw with a high-energy cutoff, which is influenced by the maximum energy of the electrons in the corona. 
Additionally, when a portion of the Compton-upscattered hard X-rays interact with the thin accretion disc, they undergo absorption and scattering, leading to the formation of a reflection component \citep{1974A&A....31..249B, 1988ApJ...335...57L}, which consists of fluorescent line emission from various elements, a soft thermal continuum, and a bump-like feature called Compton hump peaked at $\sim$ 20$-$30 keV. 
Also, there may exist absorption features in the spectra that indicate the presence of accretion disc wind comprised of neutral or highly ionized absorbers in BH-XRB systems \citep{1997xisc.conf..427E, 2002ApJ...567.1102L,2023MNRAS.520.4889P}. These features are revealed through high-resolution X-ray spectroscopy.

The study of a BH-XRB system primarily involves identifying its fundamental binary parameters and understanding its dynamic behavior throughout an outburst.
The mass and spin of the BH, the inclination of the disc with respect to the line of sight of the observer, and the distance to the system are the fundamental parameters of a BH-XRB system.
The mass of a black hole in an X-ray binary system is inferred by studying the orbital motion of the companion star, utilizing radial velocity measurements of emission lines. Apart
from this method, the mass can also be estimated using X-ray spectral modeling.

During outbursts, BH-XRB systems transition through a range of spectral states. The evolution of BH-XRBs during an outburst through different states can be visualized using their Hardness-Intensity Diagram (HID) \citep{2001ApJS..132..377H,2004MNRAS.355.1105F,2006ARA&A..44...49R,2020MNRAS.497.1197B,2022MNRAS.514.6102P}. The hardness or hardness ratio (HR) is defined as the ratio of flux in a higher-energy band to that in a lower-energy band. 
A typical canonical outburst begins with the low/hard state (LHS) and progresses to the high/soft state (HSS) through the hard and soft-intermediate states (HIMS and SIMS) \citep{2001ApJS..132..377H,2005A&A...440..207B,2022MNRAS.514.6102P}. Such outbursts exhibit a typical counter-clockwise `q'-shape in the hardness-intensity diagram \citep{2005Ap&SS.300..107H,2018Ap&SS.363..189D,2022MNRAS.514.6102P}. 
There will always be numerous candidates exhibiting unusual behavior that could lead us to discover new perspectives on the behavior of these systems. The q-shaped behavior (canonical outbursts) is not observed in all sources, and we define them as `non-canonical' outbursts. They are usually soft state dominated or have very short-lived hard state that go unnoticed by observations. The HIDs of these sources are quite different from the typical q-shaped profile and mostly replicate the HID behavior during the soft state. For example, 4U 1543-47, 4U 1630-472, and MAXI J1631-479 are soft state dominated and lack a typical hard state.

This manuscript aims to unveil the characteristics and behavior of the source 4U 1755$-$338, a quasi-persistent system \citep{2016A&A...587A..61C}, during its non-canonical outburst in 2020 \citep{2020ATel13606....1M}. 
The source 4U 1755$-$338, identified as an LMXB, harbors a black hole candidate (BHC) located in the direction of the Galactic center with $RA=17^h 58^m 40^s$, $Dec=-33^\circ 48^{'} 26^{''}.8$  (J2000). 
It was discovered by the \textit{Uhuru} satellite in 1970 \citep{1974ApJS...27...37G,1977ApJ...214..856J} and may have exhibited activity even earlier. This X-ray binary is among the earliest detections in its category. 
The source exhibits an unusually soft spectrum \citep{1977ApJ...214..856J}, and has a lower equivalent neutral hydrogen column density, $N_H$ ($\sim$ 2.2 $\pm$ 1.2 $\times$ 10$^{21}$ cm$^{-2}$) \citep{1984ApJ...281..354W}. From \textit{EXOSAT} observations, \citet{1984ApJ...283L...9W} identified three evenly spaced dips in the lightcurve of 4U 1755$-$338, indicating a binary period of approximately 4.4 hours for this system. They argued that the dips are caused by the regular occultation of the X-ray source by a thickened region of an accretion disc at the point where it is fed by the binary companion. They also found that the dips are spectrally independent and the hardness ratio remains nearly constant across the dip and non-dip regimes, suggesting a critical underabundance of heavy elements.
\citet{1984ApJ...281..354W} and \citet{1984ApJ...283L...9W}
identified 4U 1755$-$338 as a BHC based on its position in the X-ray color-color diagram.
\citet{1995MNRAS.274L..15P} observed a hard powerlaw tail in the X-ray spectrum of 4U 1755$-$338 during the \textit{TTM Galactic Centre survey} of 1989, in addition to the ultrasoft thermal component, confirming the black hole candidacy of 4U 1755$-$338. Using simultaneous \textit{Ginga} and \textit{ROSAT} observations, \citet{1995ApJ...454..463S} detected the powerlaw tail and also an iron emission line centered at  $\sim$ 6.7 keV in the spectra of the low-luminosity state (referred to as the ``low" state, which is relatively harder than the high-luminosity state, or ``high" state) of the source during 1991. The observed equivalent width of the iron line aligns with that of other LMXBs \citep{1995ApJ...454..463S}, addressing the concern regarding the under-abundance of heavy elements. In addition, they reported an $N_H$ of $\sim$ 4 $\times$ 10$^{21}$ cm$^{-2}$ \citep{1995ApJ...454..463S} to the system.

X-ray emission from this source continued until \textit{RXTE} observed a transition to quiescence in January, 1996 \citep{1996IAUC.6302....2R}. The flux decreased by a factor of 100 from the typical value. \citet{1998ApJ...496L..21W} studied optical observations during the quiescent period in 1996 to determine the nature of the companion star. Using period-mass relations and the 4.46-hour period of the system, they estimated the companion star's mass to be $\sim$0.5 $M_{\odot}$. They also suggested that if the donor is an M0V star, as implied by the period, the upper limit on the brightness of the counterpart indicates a distance greater than 4 \textit{kpc}. 
Distance estimates to 4U 1755$-$338 vary significantly, ranging from 4 to 9 kpc
\citep{1998ApJ...496L..21W}.
A study by \citet{2003ApJ...586L..71A} of the 2001 \textit{XMM-Newton} observations of 4U 1755$-$338 in its quiescent state identified a prominent X-ray jet-like feature centred on the source.  The spectral characteristics of this jet resemble those observed in other Galactic and extragalactic jets associated with accreting black holes. This detection provides additional evidence in support of the black hole candidacy of 4U 1755$-$338. \textit{Chandra} observations in 2003 unveiled the diffuse nature of this jet-like feature \citep{2005ApJ...618L..45P}. \citet{2006ApJ...641..410K} further examined \textit{XMM-Newton} observations from 2004 and verified the presence of the X-ray jet.

Recently, 4U 1755$-$338 entered into an outburst on April 2, 2020 \citep{2020ATel13606....1M}. This time, the outburst is being monitored by a range of sophisticated X-ray instruments, including \textit{Neutron Star Interior Composition Explorer (NICER)} \citep{2016SPIE.9905E..1HG} and \textit{Nuclear Spectroscopic Telescope
Array (NuSTAR)} \citep{2013ApJ...770..103H}. 
No detailed study exists in the literature on the re-brightening of 4U 1755$-$338.
In this manuscript, we perform a comprehensive spectral and timing analysis of the 2020 outburst of 4U 1755$-$338 using observations from \textit{NICER} and \textit{NuSTAR}.
The source continues the 2020 outburst at the time of writing. We performed a thorough spectral modeling throughout the entire \textit{NICER} epoch and also performed simultaneous \textit{NICER-NuSTAR} modeling for further investigation.

\section{OBSERVATIONS AND DATA REDUCTION} \label{sec:sec2}
To date, \textit{NICER} has full coverage of the 2020 outburst of 4U 1755$-$338. \textit{NuSTAR} has a single observation (ID: 90601313002) starting on MJD 58948 with an exposure time of 32,022 seconds. 
We conducted this study using all \textit{NICER} observations with exposure greater than 500 s, covering a period from 2 April 2020 (MJD 58941) to 17 August 2023 (MJD 60173) and the single \textit{NuSTAR} observation that occurred on 9 April 2020. See Table~\ref{tab:obs_list} for the details of the observations used for this study. A simultaneous \textit{NICER} observation coincided with the \textit{NuSTAR} epoch, and this combination was utilized to conduct wideband spectral modeling.
\begin{table}
    \centering
    \caption{Summary of all the observations considered for the study.}
    \setlength{\tabcolsep}{5.5pt} 
	\renewcommand{\arraystretch}{2.0} 
    \begin{tabular}{|ccc|}
    \hline
      Instrument   & Start date (MJD) & End date (MJD)  \\
    \hline
    \textit{NICER}  & 58941 & 60173 \\ 
    \hline 
    \hline
    \multicolumn{3}{|c|}{Simultaneous \textit{NICER-NuSTAR} epoch} \\
    \textit{NICER} (ID: 3201080107) & 58948.52 & 58948.65 \\
    \textit{NuSTAR} (ID: 90601313002) & 58948.32 & 58949.11  \\
    \hline
    \end{tabular}
    \label{tab:obs_list}
\end{table}

\begin{figure}
	\centering
	\includegraphics[width=0.83\textwidth]{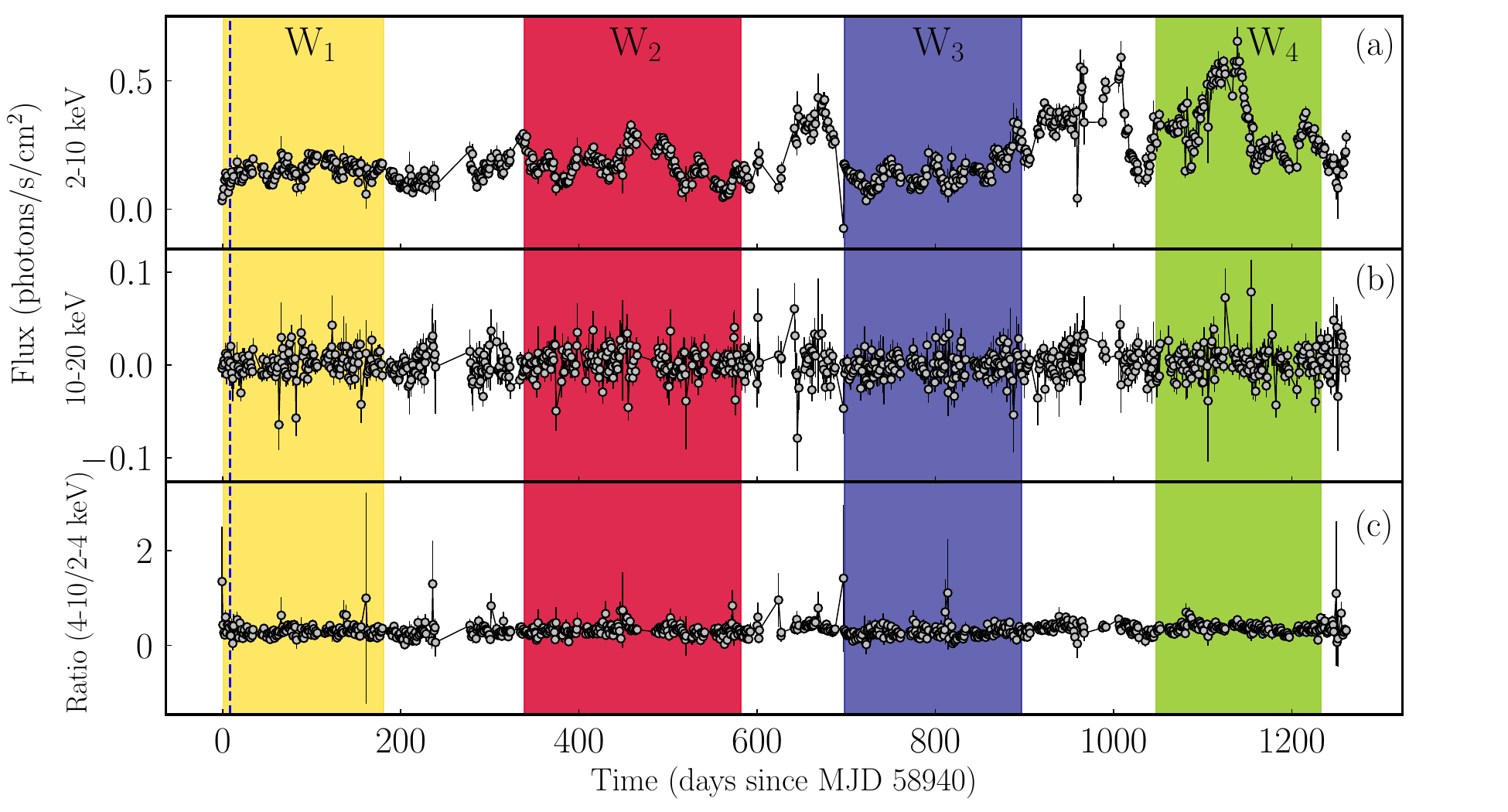}
	\caption{The \textit{MAXI/GSC} lightcurve of 4U 1755$-$338 during the 2020 outburst in the energy bands \textbf{(a)} 2$-$10 keV and \textbf{(b)} 10$-$20 keV. \textbf{(c)} The ratio of  fluxes in $4-10$ keV to $2-4$ keV defines the hardness. The four different \textit{NICER} observation windows considered for this study, spanning the outburst period, are shown in yellow, red, blue, and green colors. The blue dashed line marks the single \textit{NuSTAR} observation on day 8 of the outburst.}
	\label{fig:f1}
\end{figure}

\subsection{\textit{NICER} Data Reduction} \label{sec:nicer}

The \textit{NICER} data is reduced using \texttt{NICERDAS 11a}\footnote{\url{https://heasarc.gsfc.nasa.gov/docs/nicer/nicer_analysis.html}} in \texttt{HEASOFT v.6.32.1} with the \texttt{20221001 caldb} version. The calibration and screening of the data is taken care of by the \texttt{nicerl2} task. Scientific products, such as spectra and lightcurves are extracted using the \texttt{nicerl3} task. The latest \texttt{SCORPEON} model is used to generate the background file, producing an adjustable background model that can be fitted within the \texttt{XSPEC} \citep{1996ASPC..101...17A} spectral fitting platform along with the source spectrum. The task \texttt{nicerl3-spect}  automatically applies a systematic error of 1.5$\%$ to the data.
Optimal binning with a minimum of 10 counts is applied to all \textit{NICER} spectra using the tool \texttt{ftgrouppha}.

\subsection{\textit{NuSTAR} Data Reduction} \label{sec:nustar}
\textit{NuSTAR} comprises two focal-plane module telescopes, denoted \textit{FPMA} and \textit{FPMB}, both operating over the energy range 3 to 78 keV. The \textit{NuSTAR} observation of 4U 1755$-$338 is reduced using \texttt{NUSTARDAS\footnote{\url{https://heasarc.gsfc.nasa.gov/docs/nustar/analysis/}} pipeline v.2.1.1} and \texttt{CALDB v.20230613}. 
A circular source region with a radius of 40 pixels, centered on the position of 4U 1755$-$338, is selected, and a background region of similar size is chosen away from the source. 
These files are used to produce
various scientific outputs, including the spectrum, background, lightcurve, Auxiliary Response File (ARF), and Response Matrix File (RMF), utilizing the \texttt{NUPRODUCTS} task separately for both \textit{FPMA} and \textit{FPMB}. The spectrum is binned with a minimum of 40 counts per bin, without incorporating any systematics.

\section{Outburst profile} \label{sec:outburst}
The 2020 outburst of 4U 1755$-$338 is a low-luminosity outburst. The \textit{MAXI/GSC} (\textit{Monitor of All-sky X-ray Image/Gas Slit Camera}, \citep{2009PASJ...61..999M,2011PASJ...63S.623M}) daily lightcurve of the outburst in the energy bands, 2$-$10 keV and 10$-$20 keV are shown in Fig.~\ref{fig:f1}a and Fig.~\ref{fig:f1}b respectively. 
In this study, we designated 1 April 2020 (MJD 58940) as day 0 of the 2020 outburst, and we plotted the light curve over 200 days. The lightcurve in the 2$-$10 keV band shows an overall increase in the flux values as the outburst proceeds, with the highest observed flux of 0.6541 photons/sec/cm$^2$. In the 10$-$20 keV band, the observed flux is significantly low. Hence, we defined the hardness ratio as the ratio of flux in 4$-$10 keV to 2$-$4 keV, which is plotted in Fig.~\ref{fig:f1}c.
During the course of the outburst, \textit{NICER} provided continuous coverage of the source, spanning four observation windows (W$_1$, W$_2$, W$_3$ and W$_4$). The timeline of the \textit{NICER} observation windows considered for this study is represented by four distinct colors in Fig.~\ref{fig:f1}: yellow (W$_1$), red (W$_2$), blue (W$_3$), and green (W$_4$). 

\section{Spectral Analysis and Results} \label{sec:spec-result}
To comprehend the characteristics and behavior of 4U 1755$-$338 during the 2020 outburst, we performed a thorough spectral analysis using \textit{NICER} observations. We used the \texttt{XSPEC V12.13.1} platform for spectral modeling. The evolution of the parameter space is studied, and the HID of the source is generated during this period. 
There exists a \textit{NICER-NuSTAR} simultaneous epoch that is utilized to perform the wideband spectral modeling. Also, we attempted to calculate the fundamental binary parameters of the system from the wideband spectral modeling, including the BH mass, spin parameter, disc inclination, etc. The details of the observations used for this study are provided in Table~\ref{tab:obs_list}.

\subsection{Comprehensive Spectral Analysis} \label{sec:spec-nicer} 
\begin{figure}
	\centering
	\includegraphics[width=0.7\textwidth]{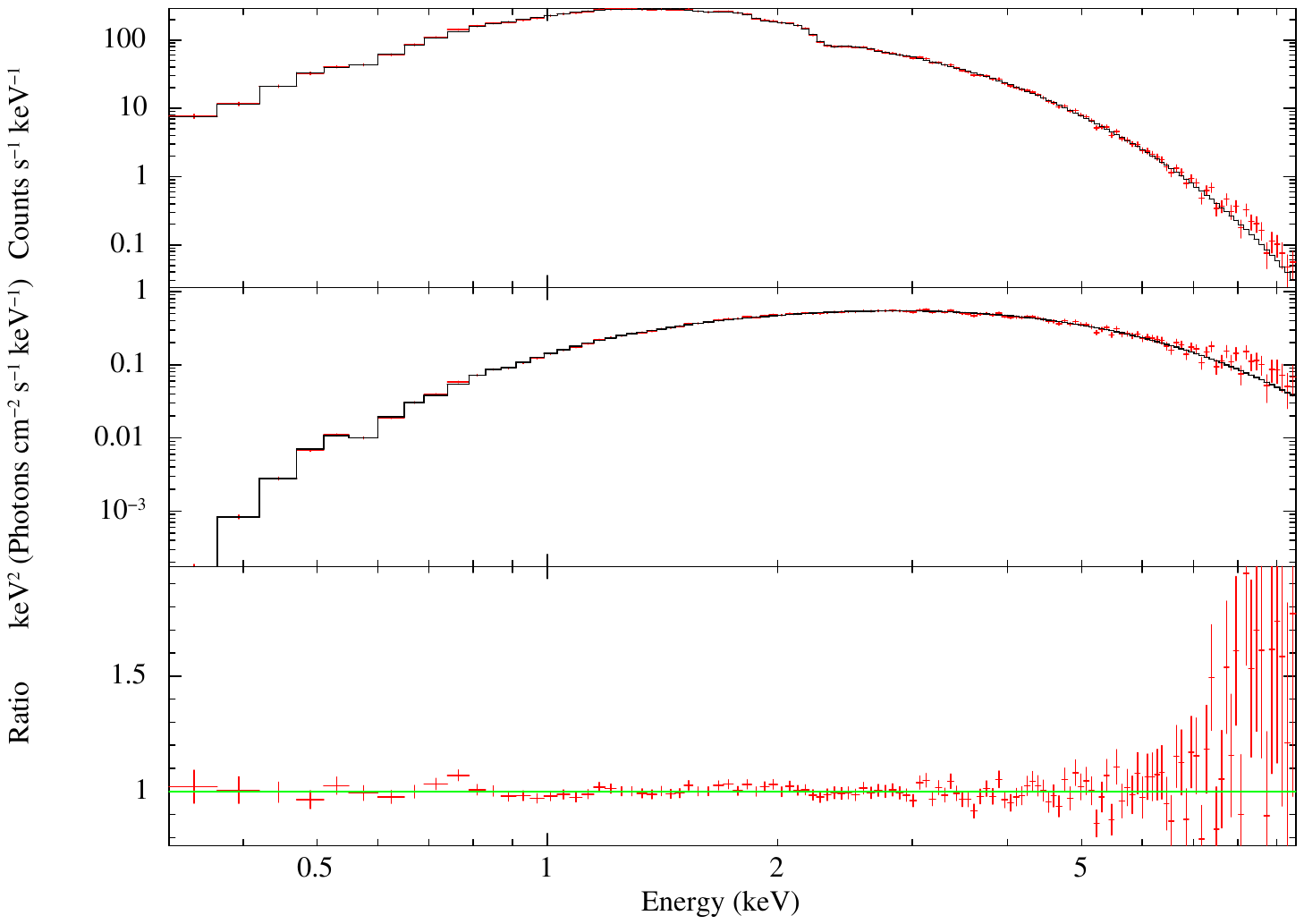} 
	\caption{Fitted spectrum of the \textit{NICER} observation (ID: 3201080122) on day 127 of the outburst using the model $tbabs \times tbpcf \times diskbb$. The top panel shows the modelled data, the middle panel represents the unfolded spectrum, and the bottom panel shows the fitting residual.  }
	\label{fig:f8}
\end{figure}

All available \textit{NICER} data with an exposure time greater than 500 seconds are used for the comprehensive analysis. 
The latest version of the \textit{NICER} software  (\texttt{NICERDAS 11a}) used for this study allows for a joint source and background spectral modeling. 
We performed a combined spectral fit for each \textit{NICER} window separately, and the Galactic $N_H$ is tied across observations of each window. Since the number of observations in W$_3$ and W$_4$ is small, we fit them together. There are three sets of combined fittings in total.
First, the data are fitted between 0.3$-$10 keV using the model $tbabs \times diskbb$ to examine the spectral behavior. The \textit{tbabs} model in \texttt{XSPEC} is used to address interstellar absorption, where the equivalent hydrogen column density, $N_H$, is provided through the solar abundance table \citep{2000ApJ...542..914W}. The \textit{diskbb} \citep{1984PASJ...36..741M} component is used to model the multicolor blackbody spectrum from the accretion disc.  
The \textit{diskbb} model has two components, $T_{in}$, the temperature of the inner accretion disc, and \textit{norm}, the normalization. 
The residual obtained after applying the above model showed a low-energy absorption feature at $<$ 1 keV. We tried several absorption models and partial covering models in XSPEC, such as \textit{phabs, pcfabs, zxipcf, tbpcf} 
\citep{2021ApJ...919...90I,2000ApJ...542..914W}, etc. to address this feature, and finally we compared the $\chi^2$ values between these models and decided to proceed with the \textit{tbpcf} model \citep{2000ApJ...542..914W}, which provided the most reasonable fit. The \textit{tbpcf} is a partial covering absorption model, which has two parameters: equivalent hydrogen column density ($N_H^{pc}$), partial covering fraction (\textit{pcf}), and the redshift parameter, $z$ ($z$ is frozen at 0). 
Hence, the final model to fit the \textit{NICER} data is chosen as $tbabs \times tbpcf \times diskbb$ (M$_1$,  hereafter). 
Fig.~\ref{fig:f8} shows a sample fit for the \textit{NICER}  observation (ID: 3201080122) on day 127 of the outburst. The $N_H$ for this observation is frozen at $0.34\times10^{22}$ cm$^{-2}$ obtained from window-wise spectral fitting as mentioned above, and the estimated spectral modeling parameters are 
$N_H^{pc}=0.24\pm{0.04}$, $pcf=0.65_{-0.05}^{+0.06}$, $T_{in}=1.12\pm{0.01}$ and \textit{diskbb norm}$= 54.87_{-1.87}^{+1.96}$. The goodness of the fit, $\chi^2_{red}=\chi^2/dof=1.1$, is determined using $\chi^2$ statistics, and the uncertainties in the parameters are quoted at 90\% confidence.
Also, the significance of \textit{tbpcf} component is verified using F-test with F statistic value $=$ 52.90 and probability $=9.46\times10^{-18}$. 

\subsection{Hardness-Intensity Diagram of 4U 1755-338} \label{sec:HID}
\begin{figure*}   
	\centering
		\includegraphics[width=0.49\textwidth]{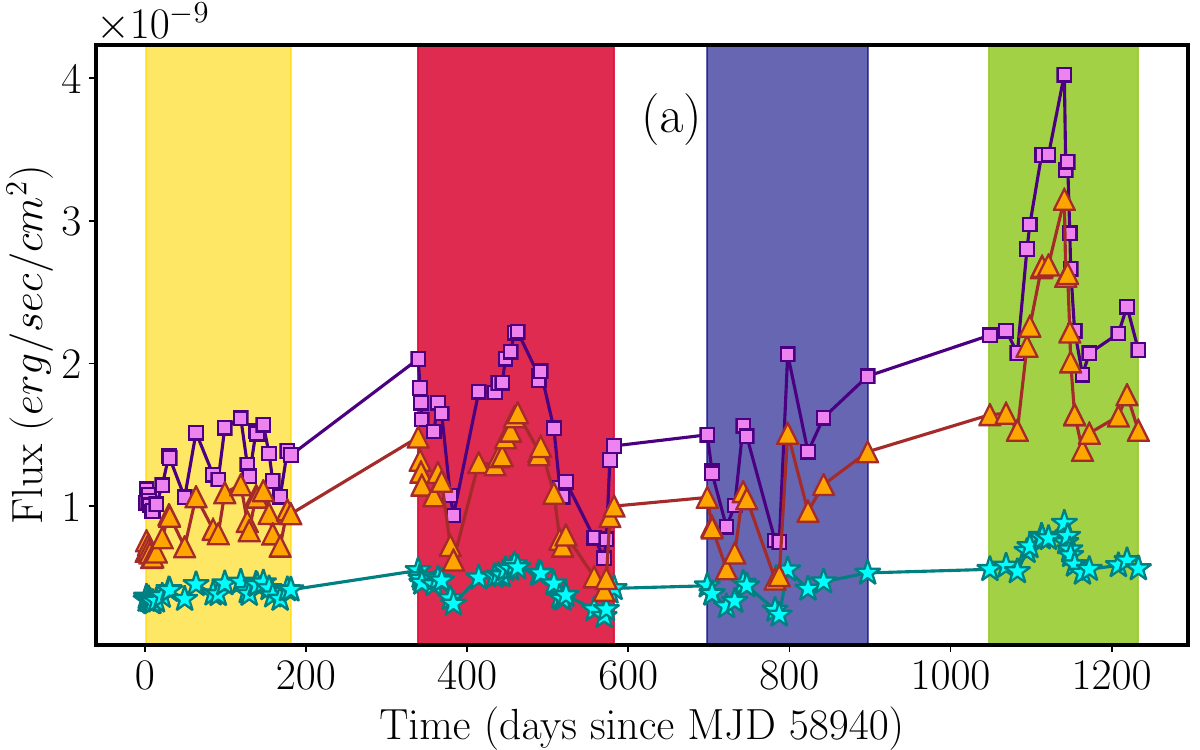}
		\includegraphics[width=0.46\textwidth]{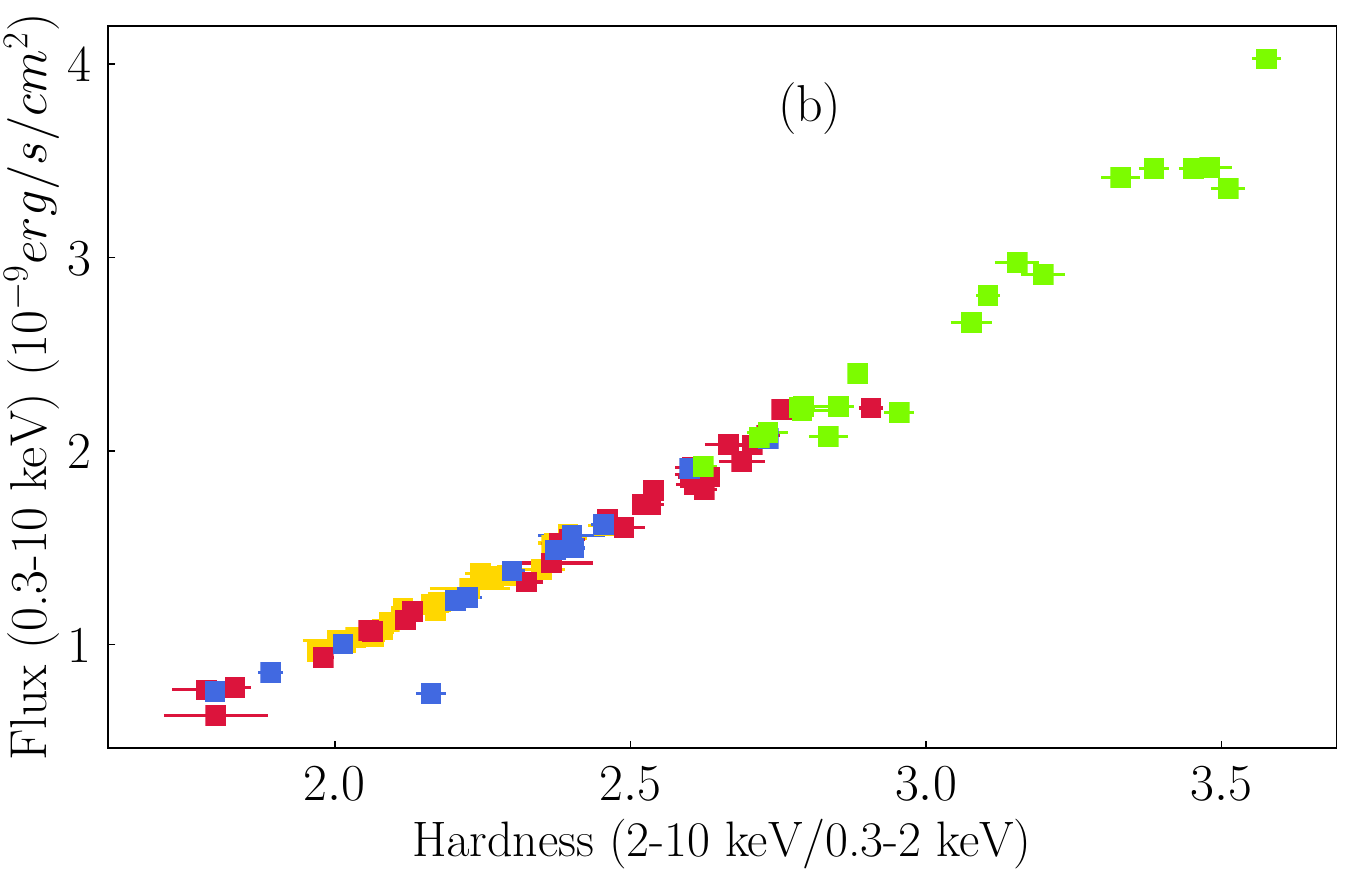}
	\caption{ \textbf{(a)} The \textit{NICER} lightcurve of 2020 outburst of 4U 1755$-$338 in different energy bands, 0.3$-$2 keV, 2$-$10 keV, and 0.3$-$10 keV represented in cyan colored stars, orange triangles, and violet squares respectively. The four windows of \textit{NICER} observation are marked with their respective colors as previously discussed. \textbf{(b)} Hardness-Intensity Diagram of 4U 1755$-$338 using \textit{NICER} observations. The colors yellow, red, blue, and green are used to distinguish between the four windows of \textit{NICER} observation. The uncertainties are within the 90 per cent confidence range. 
    }
	\label{fig:f56}
\end{figure*}

We estimated the flux from the spectral fitting using the model M$_1$ in the energy bands, 0.3$-$2 keV, 2$-$10 keV, and 0.3$-$10 keV, from all the \textit{NICER} observations and are plotted in Fig.~\ref{fig:f56}a using cyan stars, orange triangles, and violet squares, respectively. The four \textit{NICER} observation windows (W$_1$ to W$_4$) are marked in yellow, red, blue, and green colors in their chronological order, as discussed in \S\ref{sec:outburst}.
We now define the \textit{hardness} (HR) as the ratio of flux in 2$-$10 keV to the flux in 0.3$-$2 keV. HR is plotted against the total flux (in 0.3$-$10 keV) to generate the HID (Fig.~\ref{fig:f56}b). The colors yellow, red, blue, and green are used to distinguish between the four \textit{NICER} windows W$_1$ through W$_4$ in the HID. 
Observations during W$_4$ are the hardest among the \textit{NICER} windows, with the value of HR ranging between 2.8$-$3.6, while for W$_1$ to W$_3$, the HR varies between 1.8$-$2.9. 

\subsection{Evolution of Spectral Parameters} \label{sec:spec-param}
\begin{figure*}
	\centering	\includegraphics[width=0.49\textwidth]{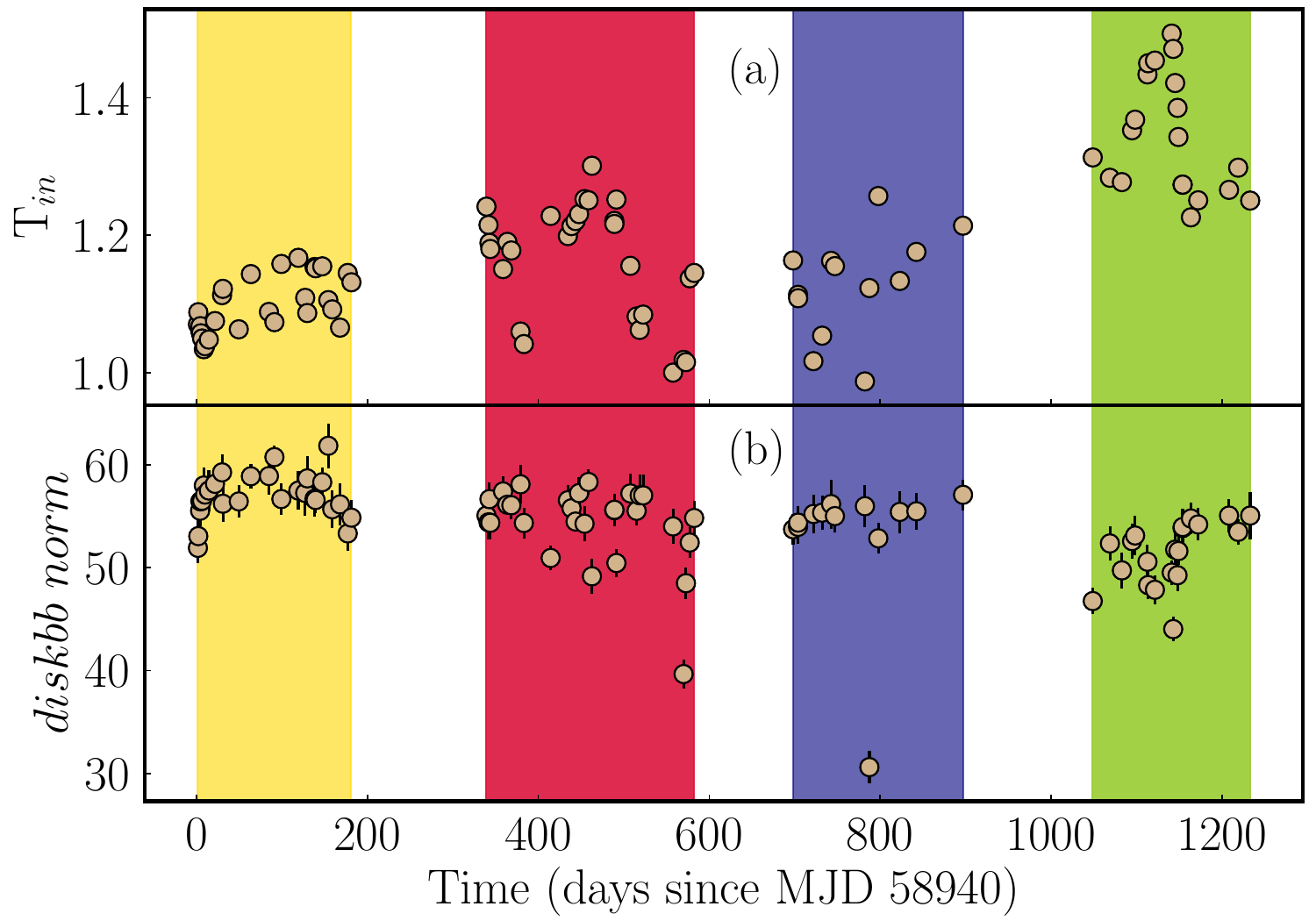}    \includegraphics[width=0.49\textwidth]{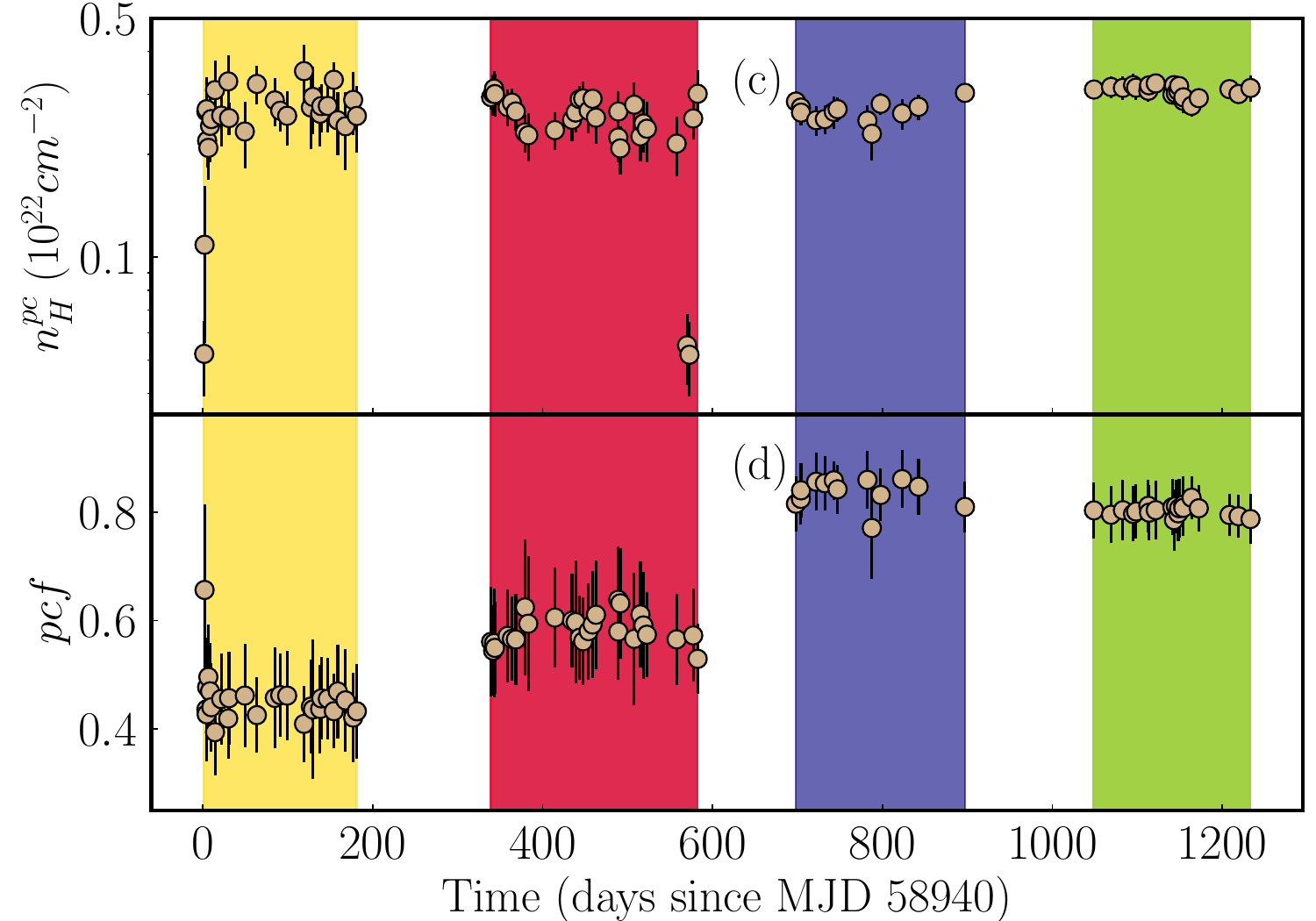}
	\caption{Evolution of the \textit{diskbb} \textbf{(a, b)} and \textit{tbpcf} \textbf{(c, d)} parameters obtained from the spectral modeling of \textit{NICER} observations using the model M$_1$. Panel \textbf{(a)}, the inner disc temperature, $T_{in}$ \textbf{(b)} the \textit{diskbb norm}, \textbf{(c)} the equivalent hydrogen column density, $N_H^{pc}$ and \textbf{(d)} the partial covering fraction, \textit{pcf}. The error bars represent 90$\%$ uncertainty.}
	\label{fig:f23}
\end{figure*}

The model parameters obtained from the spectral modeling of the \textit{NICER} data using the model M$_1$ are estimated with an uncertainty of 90$\%$ and their evolution is plotted in Fig.~\ref{fig:f23}. The four windows of the \textit{NICER} observations are highlighted using their respective colors as discussed in \S\ref{sec:outburst}. 
The average value of the line-of-sight absorption column density ($N_H$) obtained from the combined spectral fitting of three sets of \textit{NICER}data is $0.34\pm0.01 \times$ 10$^{22}$ cm$^{-2}$.
Fig.~\ref{fig:f23}a and~\ref{fig:f23}b show the evolution of the \textit{diskbb} parameters. The value of $T_{in}$ shows a gradual but slow increase from W$_1$ through W$_3$ and then shows a remarkable increase in W$_4$ according to the source flux, reaching the maximum value of 1.48 kev and then returning to the starting value of W$_4$. 
The \textit{diskbb norm} remains consistent throughout the outburst, with a value between $\sim$ 50-60
except for a few observations. 

Evolution of the \textit{tbpcf} model parameters is shown in Fig.~\ref{fig:f23}c and~\ref{fig:f23}d. The value of $N_H^{pc}$ varies between $ (0.21-0.35) \times$ 10$^{22}$ cm$^{-2}$ (except for few observations) throughout the outburst. The \textit{pcf} is $\sim 0.40$ at the beginning of the outburst, and it evolves to a higher value of $\sim$0.86 as the outburst progresses.  



\begin{figure}   
	\centering
		\includegraphics[width=0.6\textwidth]{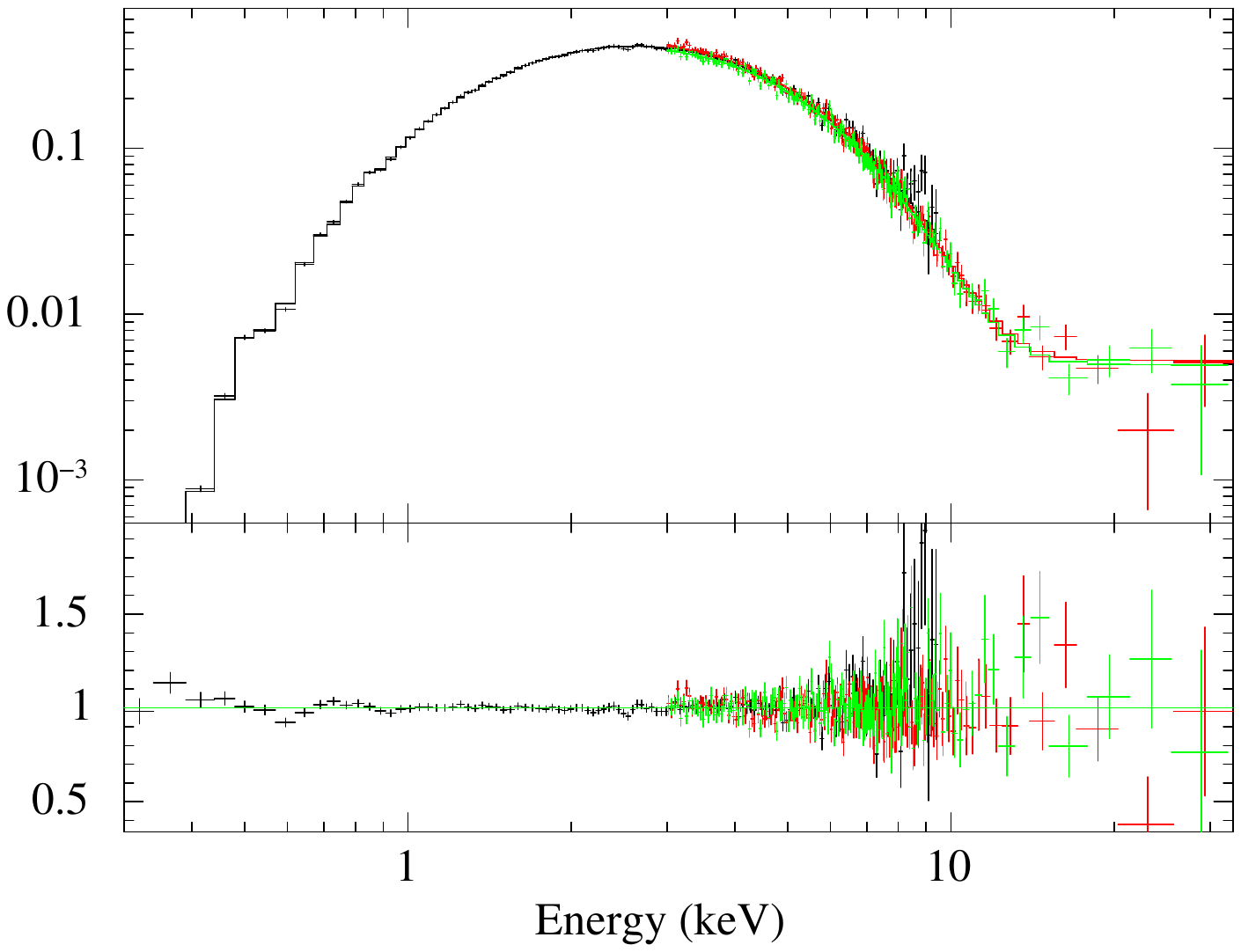}
	\caption{Wideband \textit{NICER-NuSTAR} fitted spectra using the model  $tbabs \times tbpcf \times  simpl  \times kerrbb \times constant$ for $i=80^{\circ}$ and $D=8$ $kpc$. 
    The \textit{NICER} spectrum is shown in black color, and \textit{FPMA} and \textit{FPMB} spectra of \textit{NuSTAR} are shown in red and green colors, respectively. }
	\label{fig:f910}
\end{figure}

\subsection{Wideband NICER-NuSTAR modeling} \label{sec:spec-simult}
On day 8 of the 2020 outburst, the source is observed by the X-ray instruments \textit{NICER} and \textit{NuSTAR}.
We performed a simultaneous wideband spectral analysis using the \textit{NICER-NuSTAR} data on that day. The timing of their occurrence is shown in Table~\ref{tab:obs_list}. The \textit{NuSTAR} data is time-filtered according to the start and stop times of the \textit{NICER} data. The \textit{NICER} and \textit{NuSTAR} data are modelled between 0.3$-$9 keV and 3$-$34 keV, respectively, as significant data are not available beyond this energy range. 

The primary objective of the wideband spectral study is to estimate the fundamental properties of the system, such as the mass and spin of the BH, the inclination and distance of the system, etc., using appropriate models. Usually, spin estimates from continuum fitting are reliable for sources with known mass, inclination, and distance. However, in \citet{2016ApJ...821L...6P}, the spin and inclination were first determined through relativistic reflection modeling and then adopted as input parameters for the continuum fitting model to derive mass and distance.
In our case, the spectra are devoid of any reflection signatures, and therefore, the above methods will not be applicable. However, since the data are statistically good, we proceed with the continuum fitting method to estimate the fundamental parameters.
The soft emission from the accretion disc is modelled using the \texttt{XSPEC} model, \textit{kerrbb}. The interstellar absorption and partial covering absorption models are used as before. In the wideband spectrum, there exists a high-energy tail, which is modelled using \textit{simpl} along with \textit{kerrbb}. 
The \textit{simpl} is an \texttt{XSPEC} model with parameters describing the \textit{spectral index} ($\Gamma$), and fraction of seed photons scattered into the powerlaw component ($FracSctr$). Note that the data can also be modelled using a \textit{powerlaw} together with \textit{kerrbb}. However, dominance of the \textit{powerlaw} model over the blackbody component at low energies is unphysical. Therefore, we adopt the \textit{simpl} model instead.
Hence, we have the model combination, $tbabs \times tbpcf \times simpl \times kerrbb \times constant$ (hereafter M$_2$).  The \textit{constant} is used to address the calibration differences between the instruments. 
The model \textit{kerrbb} \citep{2005ApJS..157..335L} in \texttt{XSPEC} produces a multicolor blackbody spectrum of a geometrically thin, steady state, general relativistic accretion disc around a rotating black hole.
The major components of this model are the \textit{$\eta$}, the ratio of the disc power produced by a torque at the inner disc to that arising from the accretion, 
\textit{a}, the specific angular momentum of the BH (dimensionless black hole spin), \textit{i}, the inclination angle of the disc rotation axis to the line of sight in \textit{degrees}, $M$, the mass of the black hole in units of solar mass ($M_{\odot}$), $\dot M$, the effective mass accretion rate in units of $10^{18}$ $g/s$, and $D$, the distance of the black hole from the observer in $kpc$ and spectral hardening factor ($f_{col}$). We set up a standard Keplerian disc with zero torque at the inner boundary (\textit{$\eta$}$=0$). All model parameters are kept free and the \texttt{steppar} command is used to obtain the best-fit model parameters.

\begin{figure}
	\centering
	\includegraphics[width=0.49\textwidth]{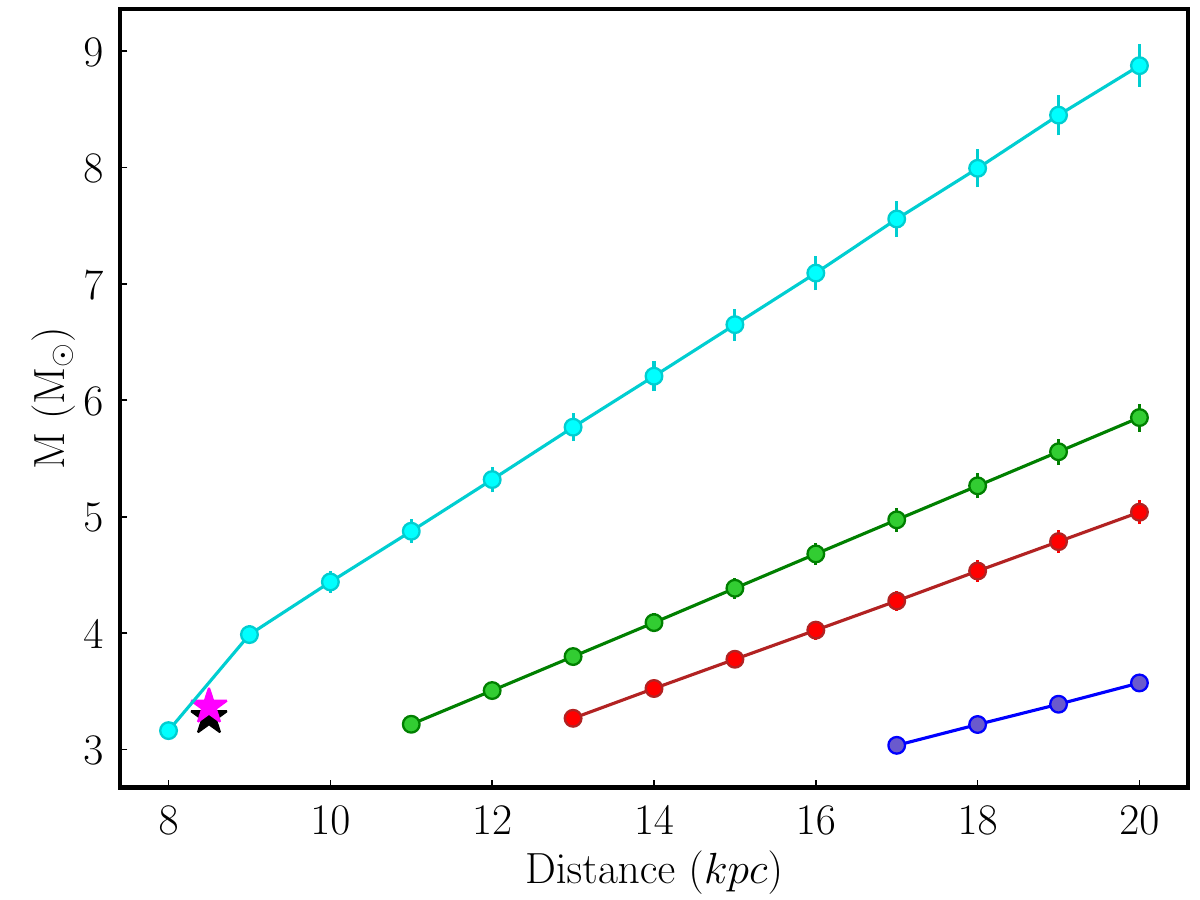}
	\caption{The evolution of the black hole mass as a function of distance, for different inclination angles, obtained from the simultaneous \textit{NICER-NuSTAR} fitting using the model $tbabs \times tbpcf \times simpl \times kerrbb \times constant$ (model M$_2$), is represented by the colors, blue (50$^{\circ}$), red (60$^{\circ}$), green (70$^{\circ}$), and cyan (80$^{\circ}$) respectively. The black and magenta star symbols represent the corresponding parameters obtained from model M$_2$ and M$_3$, respectively using MCMC methods for parameter estimation, keeping mass and inclination as free parameters for a fixed distance of $8.5$ $kpc$ (see Table~\ref{tab:sim_param}).}
	\label{fig:f12}
\end{figure}
A sample fitted spectrum for $i=80^{\circ}$ and $D=8$ $kpc$ modelled using M$_2$ is shown in Fig.~\ref{fig:f910}. The $\chi^2_{red}$ of the fitting is $1.06$ and the best-fit parameters with 90$\%$ uncertainty are as follows: $N_H=0.44\pm0.01\times$10$^{22}$ cm$^{-2}$, $N_H^{pc}=1.10_{-0.28}^{+0.34}\times$10$^{22}$ cm$^{-2}$, \textit{pcf}$=0.25_{-0.03}^{+0.05}$, \textit{a}$=0.40$ (unable to constrain the uncertainty), \textit{M}$=3.16\pm{0.07}$ $M_{\odot}$,  $\dot M$ $=0.356_{-0.004}^{+0.005} \times$ 10$^{18}$ $g/s$, $\Gamma=2.02_{-0.38}^{+0.41} $,  $F racSctr=0.010_{-0.002}^{+0.003}$ and $f_{col}=1.8$.
We model the wideband data using model-M$_2$ and attempt to estimate all parameters. However, estimating the uncertainty of many parameters, particularly the inclination ($i$) and distance ($D$), using the \textit{error} command in \texttt{XSPEC} was challenging. Hence, we kept $i$ and $D$ frozen at specific values to study the evolution of the mass. 
We assumed a set of fixed inclinations (in an interval of $10^{\circ}$) and distances (in an interval of 1 $kpc$), then obtained the parameter space where a valid BH mass exists.
Then, all other parameters, particularly $M$ and $a$, are estimated in each case. 
The variation in mass with distance is examined and illustrated in Fig.~\ref{fig:f12}. The error bars represent 90$\%$ uncertainty. Different colors denote various inclination angles: blue, red, green, and cyan correspond to 50$^{\circ}$, 60$^{\circ}$, 70$^{\circ}$, and 80$^{\circ}$ respectively. 
At lower angles of inclination, an accepted black hole mass (i.e., $M > 3 M_{\odot}$) is observed at greater distances; for instance, according to M$_2$, at an inclination of 50$^{\circ}$, the estimated distance to the source should not be less than 17 $kpc$. Similarly, for higher inclination values, the likelihood of it being a closer system increases. The evolution of the spin parameter, $a$, with distance is also investigated for different inclination angles. We found that the value of $a$ appears to vary between $0.3$ and $0.5$ in model M$_2$, but we cannot constrain its uncertainty in all cases. 

Now, all the parameters of M$_2$ are kept free to estimate the fundamental quantities of the system. We could not constrain $f_{col}$, and assume $f_{col} = 1.8$. 
Since M$_2$ has a large number of parameters, the standard \textit{error} command in \texttt{XSPEC} does not provide a reliable estimate of the parameters.
Therefore, we employ Monte Carlo Markov chains (MCMC)\footnote{\url{https://heasarc.gsfc.nasa.gov/xanadu/xspec/manual/node43.html}} to obtain a robust estimate of the uncertainty of the parameters.    
 The best-fit value of the source distance (D) is 8.5 $kpc$; however, we could not estimate its uncertainty. Hence, we kept D frozen at 8.5 $kpc$. The source lies in the direction of the Galactic Center, and the distance estimated from spectral modeling of the high-energy tail using a \textit{powerlaw} component is $8.6_{-4.39}^{+5.10}$ $kpc$. These considerations together justify this approach. Also, \textit{tbpcf} column density ($N_H^{pc}$) is fixed at its best-fit value of 1.3, as it could not be well constrained. We used the Goodman-Weare algorithm \citep{2010CAMCS...5...65G} with 70 walkers, a total chain length of 10,00,000, and burned the first 300000 steps of the chain. The results are tabulated in Table~\ref{tab:sim_param}.

To understand the choice of $f_{col}$ parameter in $M_2$, we replace \textit{kerrbb} with \textit{bhspec} model \citep{2006ApJS..164..530D} where the color correction factor is implemented using more realistic atmospheric scattering calculations. Therefore, we performed simultaneous spectral fitting using the model, \textit{tbabs $\times$ tbpcf $\times$ simpl $\times$ bhspec $\times$ constant} ($M_{3}$ hereafter). We adopted ‘\textit{bhspec-1}’, one of the two available versions of the model, as the model M$_2$ using MCMC calculation favors a black hole spin less than 0.8. The \textit{bhspec} normalization was calculated and fixed to the value corresponding to a source distance of $D = 8.5$ $kpc$, since these two parameters are correlated. The MCMC chains of length 1000000 are created in this case using Goodman-Weare algorithm with 44 walkers, and burned the first 100000 steps.
We found that the black hole mass, spin parameter, mass accretion rate, and inclination obtained from M$_2$ and M$_3$ models are broadly consistent (See Table~\ref{tab:sim_param}). However, the spin parameter shows a marginal discrepancy between the two models. Allowing the distance to vary resulted in poorly constrained uncertainties; therefore, we chose to fix $D$ at $8.5$ $kpc$. Using the estimated source mass and spin parameters provided in Table~\ref{tab:sim_param} for the M$_2$ model, and following the methodology outlined by \citet{1972ApJ...178..347B}, we calculate the accretion efficiency factor. The derived $\dot{M}$ and calculated accretion efficiency yield a source luminosity of $0.049$ in units of the Eddington luminosity ($L_{Edd}$), and which is comparable to the value obtained from M$_3$ (see Table~\ref{tab:sim_param}).
The BH mass values estimated from M$_2$ and M$_3$ are shown (in black and magenta star symbols, respectively) in Fig.~\ref{fig:f12} for comparison, along with their distance and inclination information.

\definecolor{silver}{rgb}{0.75, 0.75, 0.75}
\begin{table}
    \centering
         \caption{The best-fit parameters obtained from the simultaneous modeling of \textit{NICER-NuSTAR} spectra using the models M$_2$ and M$_{3}$ (see text for definition). The uncertainties of all the parameters are estimated at the 90$\%$ confidence level.} 
    \setlength{\tabcolsep}{2.1pt}
	\renewcommand{\arraystretch}{1.6} 
    \begin{tabular}{|ccccccccccccc|}
    \hline
          Model & $N_H$ & $N_H^{pc}$ & \textit{pcf} & \textit{D} & \textit{i} & \textit{a} & \textit{M} & $\dot M$ & $L/L_{Edd}$ & $\Gamma $ & $\chi^2_{red}$ &\\
         & (10$^{22}$ cm$^{-2}$) & (10$^{22}$ cm$^{-2}$)&  & (\textit{kpc}) & (\textit{degree}) &  & ($M_{\odot}$) & (10$^{18}$ $g/s$)& &  &  &\\

         \hline    
         M$_2$ & $0.443_{-0.007}^{+0.006}$ & $1.3^f$ 
         & $0.27_{-0.05}^{+0.03}$ & $8.5^f$ & $75.25_{-4.68}^{+5.59}$ & $0.50_{-0.43}^{+0.19}$
        & $3.28_{-1.1}^{+1.7}$ & $0.28_{-0.07}^{+0.19}$ & $-$ & $2.0_{-0.37}^{+0.43}$ & $1.060$ &\\ 

         M$_{3}$ & $0.44_{-0.04}^{+0.02}$ & $0.65_{-0.19}^{+0.13}$ 
         & $0.43_{-0.07}^{+0.10}$ & $8.5^{*f}$ & $67.44_{-3.03}^{+9.75}$ & $0.78_{-0.14}^{+0.02}$
        & $3.37_{-1.04}^{+0.45}$ & $-$&  $0.04_{-0.002}^{+0.02}$  & $2.90_{-0.26}^{+0.45}$ & $1.059$ &\\ 

\hline 
    \end{tabular}
    \label{tab:sim_param}
    \begin{tablenotes}
			\item $^{f}$ Frozen parameter. $^*$ Normalization of \textit{bhspec} is calculated by assuming a constant distance of 8.5 $kpc$.\\
	\end{tablenotes}  
\end{table} 

\begin{figure}
	\centering
	\includegraphics[width=0.5\textwidth]{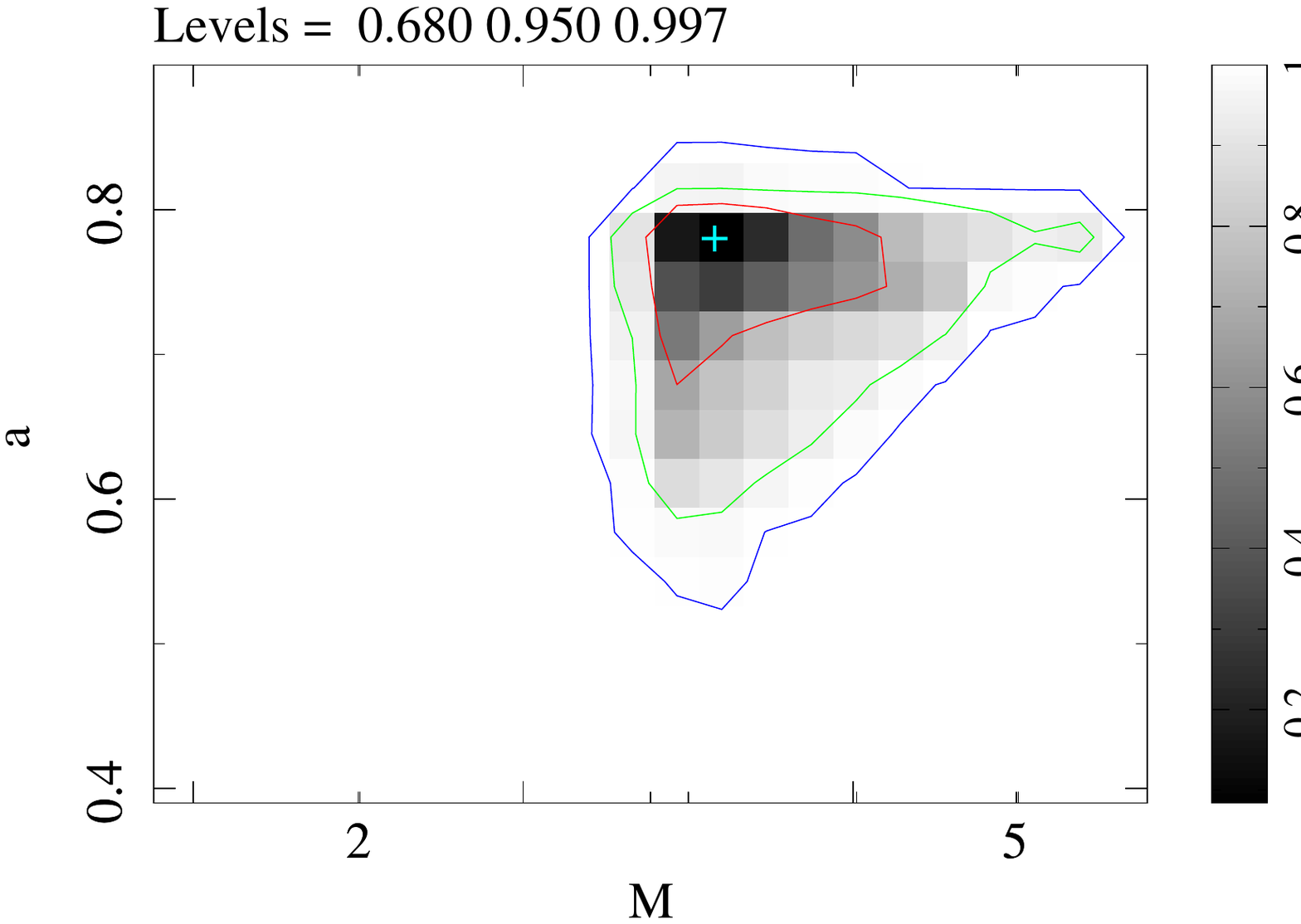}       
	\caption{The MCMC integrated probability contours of mass versus spin parameter obtained from the simultaneous \textit{NICER-NuSTAR} data modelled with \textit{tbabs $\times$ tbpcf $\times$ simpl $\times$ bhspec $\times$ constant}. The contours are plotted
for the 68$\%$ (red in color), 95$\%$ (green in color), and 99.7$\%$ (blue in color) confidence intervals, showing the range of the BH mass (in $M_{\odot}$) and spin values.}
	\label{fig:f11}
\end{figure}


\section{Timing analysis and Results} \label{sec:timing}
\begin{figure}
	\centering
	\includegraphics[width=0.5\textwidth]{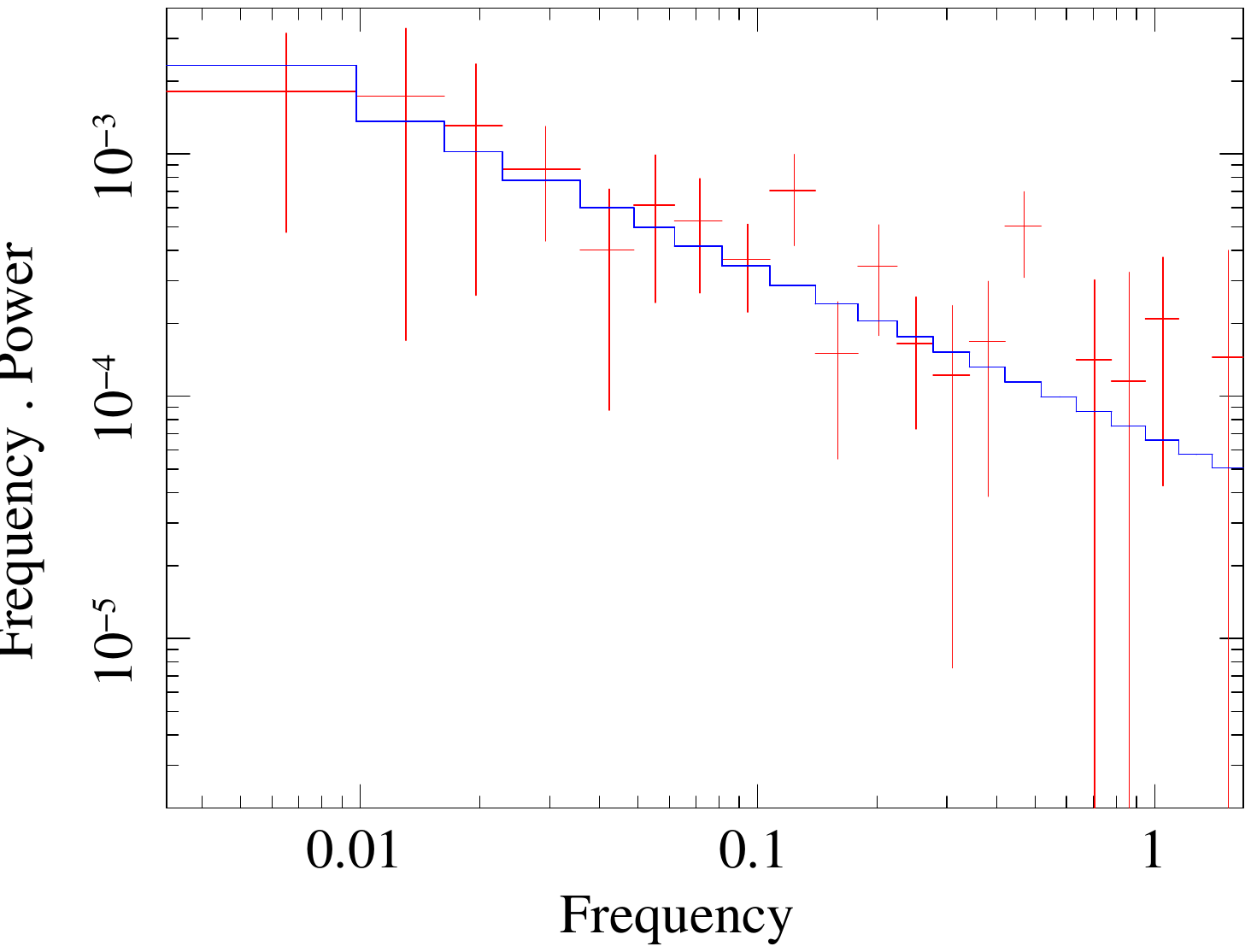}
	\caption{The power density spectrum (red in color) of the \textit{NICER} observation (ID: 6201080115) on day 1149 in 0.3$-$10 keV band fitted using a powerlaw (blue in color) with an estimated total RMS of 10.6 $\pm$ 2.7$\%$. 
 }
	\label{fig:f14}
\end{figure}

The temporal analysis uses all the \textit{NICER} data of 4U 1755$-$338 of this outburst. Lightcurves are generated in three energy bands, 0.3$-$2 keV, 2$-$10 keV and 0.3$-$10 keV (total energy), with a bin size of 0.003 seconds. 
Poison noise subtracted PDS (Power Density Spectra) are generated and are modelled in \texttt{XSPEC}. It is observed that the PDS is predominantly dominated by noise, with only a few observations providing significant data, particularly during the hardest stage of the outburst, between days 1113$-$1149. In such cases, the PDS is almost flat and can be fitted using a \textit{powerlaw} or a combination of \textit{powerlaw} and a broad \texttt{Lorentzian}. No QPO (Quasi-periodic oscillation) features are observed in any of the observations. The total RMS (in percentage) is estimated between 0.001$-$1 Hz, and the value varies between 1.6$-$2$\%$ from the total energy lightcurves. However, the observation on day 1149 shows a higher RMS (10.6 $\pm$ 2.7$\%$). The fitted power density spectrum for that particular observation is shown in Fig.~\ref{fig:f14}.
In the other two energy bands (0.3-2 keV and 2-10 keV), we were unable to estimate the RMS uncertainty.

\section{Discussion} \label{sec:discussion}
In this work, we performed a comprehensive spectro-temporal analysis of the 2020 outburst of 4U 1755$-$338 using X-ray data from two instruments, \textit{NICER} and \textit{NuSTAR}. We considered all available \textit{NICER} observations till 17 August, 2023 (with exposure greater than 500 seconds). 
This outburst comes after a long quiescence of 25 years, and there has been a lack of studies to determine the fundamental characteristics of the system, such as its distance, inclination, mass of the black hole, etc.

\begin{figure}[h]
	\centering
	\includegraphics[width=0.6\textwidth]{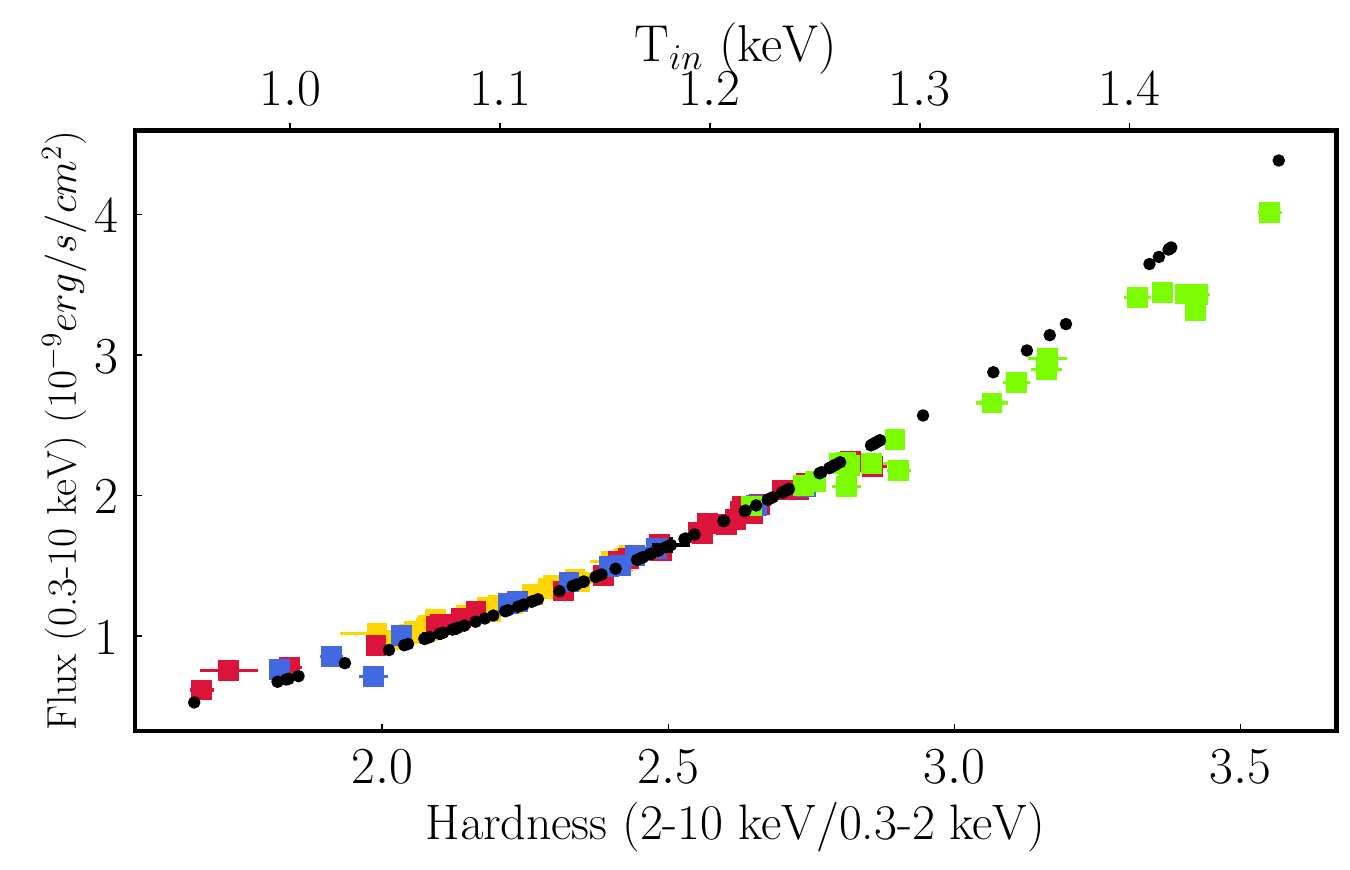}
	\caption{The hardness-intensity diagram produced for a constant \textit{diskbb norm} over the entire outburst. The colors, yellow, red, blue, and green are used to distinguish between the four windows of \textit{NICER} observation. The inner disc temperature for individual observations is marked on the top horizontal axis. Black dots represent $T_{in}^4$.  
    The uncertainties are within the 90$\%$ confidence range.}
	\label{fig:f15}
\end{figure}

First, we performed a comprehensive spectral analysis using \textit{NICER} observations of the outburst. The nature of the source flux evolution (Fig.~\ref{fig:f56}a) exhibits an overall gradual increase as the outburst progresses. For each observation window (W$_1$$-$W$_4$), the source flux shows many small flare-like signatures. This trend is more prominent in W$_4$, where the peak flux is almost twice the starting flux. The HID (Fig.~\ref{fig:f56}b) of 4U 1755$-$338 is linear in nature. The correlated evolution of flux and HR throughout the outburst suggests the thermal disc origin of the source flux. 
It is evident that the behavior of the HID does not follow the standard canonical outburst pattern with the counterclockwise q-shaped profile; rather it is mostly confined to the nature of HID in the soft state.  
To confirm this nature, we created a test case in which the \textit{diskbb norm} of all the \textit{NICER} observations is frozen to its average value and estimated the flux in the energy bands 0.3$-$2 keV, 2$-$10 keV and 0.3$-$10 keV and an HID is plotted with the same definition of hardness as Fig.~\ref{fig:f56}b. This new HID is over-plotted with the $T_{in}$ measured in this case (Fig.~\ref{fig:f15}). We used the same color code for HID to distinguish between the \textit{NICER} observation windows, as we did in \S\ref{sec:HID}, and used black dots to represent $T_{in}^4$ the evolution. The behavior of HID is consistent with the standard soft state behavior of $L\propto T^4$ (bolometric luminosity and inner disc temperature of \textit{diskbb}). Hence, the 2020 outburst of 4U 1755$-$338 can be considered a non-canonical outburst.
There are several other sources that belong to this category, for example,  4U 1630$-$472 \citep{2005ApJ...630..413T,2020MNRAS.497.1197B}, MAXI J1631$-$479 \citep{2021MNRAS.505.1213R}, 4U 1543$-$47, etc. All of these sources are predominantly observed in the soft state. However, non-canonical sources do not exhibit uniform behavior. It is not necessary for them to be in the soft state only. For example, the 2019 outburst of MAXI J1631$-$479 began with a hard state and later transitioned to HIMS \citep{2021MNRAS.505.1213R}.

The evolution of the best-fit parameters obtained from the spectral analysis revealed that the inner disc temperature $T_{in}$ shows a gradual overall increase throughout the outburst (Fig.~\ref{fig:f23}a) and reaches the highest value of 1.49 keV after 1141 days from the start of the outburst. The behavior of $T_{in}$ mimics the source flux evolution (Fig.~\ref{fig:f56}a) discussed above, which is also an indication of the thermal disc origin of the flux. 
The value of the \textit{diskbb norm} (Fig.~\ref{fig:f23}b) suggests that the inner edge of the accretion disc remains at a fixed position.

The HSS spectra are ideal for observing the signatures of an accretion disc wind, if present \citep{2002ApJ...567.1102L,2008ApJ...680.1359M,2009Natur.458..481N}. The extensive study of the entire outburst revealed an evolving disc wind of neutral gas that provides an additional column density for X-ray absorption. The value of $N_H^{pc}$ varies between ($0.21-0.35$)$\times$ 10$^{22}$ cm$^{-2}$, suggesting a cloud of more or less constant density. It is observed that the cloud partially (\textit{pcf} $\sim 45\%$ in  Fig.~\ref{fig:f23}d) covers the emission from the system at the beginning. Eventually, the size of the cloud increases to higher values ($\sim$ $60$\% coverage in W$_2$ and $\sim$ 77$-$85 \% in W$_3$ and W$_4$).  
This trend suggests a growing cloud in the system.
However, no features are observed in the spectra that confirm the presence of significant ionized winds. 
Furthermore, there is no evidence of disc reflection in the wideband spectra of 4U 1755$-$338 during the 2020 outburst. The fluorescent line, believed to originate from the reflection of hard X-rays off the inner disc, was absent from the spectra. This is expected as the source spectra are thermal disc dominated. These broad components are either not present or not clearly visible in the spectra. Partial-covering models can occasionally be used to fit such components \citep{2006ESASP.604..463F}. 
However, the presence of an iron emission line centered at $\sim$6.7 keV has been reported during the 1991 outburst of 4U 1755$-$338 \citep{1995ApJ...454..463S}.

A wideband spectral study was carried out to estimate the fundamental properties of 4U 1755$-$338, such as the mass and spin of the BH, the inclination of the system, etc. 
The study evaluates the system using two types of model approximations: M$_2$ (\textit{kerrbb}- based) and M$_3$ (\textit{bhspec}- based). The evolution of the BH mass is studied through model M$_2$ by fixing the distance and inclination of the system at specific values (Fig.~\ref{fig:f12}), and we could constrain them using the standard \texttt{XSPEC} \textit{error} command. 
In addition, we estimated all the fundamental parameters of the system using both M$_2$ and M$_3$ via the MCMC method, and found that the results are broadly consistent (Table~\ref{tab:sim_param}) between the two model approximations.

Since 4U 1755$-$338 is a low-luminosity system ($L/L_{Edd}=0.04$), disc self-irradiation and returning radiation effects are negligible. In this scenario, \textit{bhspec} modeling becomes preferable over \textit{kerrbb} because it realistically handles the vertical structure of the disc by solving full atmospheric radiative transfer (\texttt{TLUSTY}-based) for each disc annulus, calculating exact limb darkening. \textit{kerrbb}'s fixed $f_{col}$ across the entire disc does not capture how darkening varies with angle and energy at high inclinations. It may lead to overestimation of the edge brightness. However, \textit{bhspec} calculates the exact darkening for each disc position and photon energy using full atmospheric physics and relativistic ray-tracing, which is critical for estimating the spin parameter from the innermost disc emission \citep{2006ApJS..164..530D,2025ApJ...981L..15Z}. The results obtained from the \textit{bhspec} modeling (M$_3$) suggest that 4U 1755$-$338 contains a moderately spinning ($a=0.78$) BH of and mass $3.37_{-1.04}^{+0.45}$ $M_{\odot}$ located at a distance of 8.5 \textit{kpc}. The effective mass accretion rate ($\dot M$ from M$_2$) of the disc was found to be low, $0.28_{-0.07}^{+0.19}$ $\times 10^{18}$ $g/s$. 
The integrated probability contour plots of mass versus spin parameter obtained from model M$_3$ using MCMC calculation are shown in {Fig.~\ref{fig:f11}}. The colored contours represent the integrated probability for different ranges of mass and distance values; the red contour encloses 68$\%$ of the probability, the green contour encloses 95$\%$ of the probability, and the blue contour encloses 99.7$\%$ of the probability. The contour plot shows a correlation between mass and the spin parameter.

In addition, we modelled the soft disc component using the \texttt{XSPEC} model  \textit{diskpbb}, which is a variant of the \textit{diskbb} model, providing a radial dependence of the disc temperature, $T(r) \propto r^{-p}$ \citep{2005ApJ...631.1062K}. The exponent $p$ is a measurable parameter in the model whose value is 0.75 for a standard disc and less than 0.75 if advection exists. In the case of 4U 1755$-$338, we obtained $p=0.65_{-0.04}^{+0.05}$. The timing analysis reports the lack of strong variability in the PDS of 4U 1755$-$338, also, the low RMS indicates that the disc is stable.
The high value of $\Gamma$, the hardness ratio, the absence of reflection signatures, and the lack of strong variability in the PDS suggest that the source is predominantly in the soft state throughout the 2020 outburst, similar to its previous outbursts \citep{1977ApJ...214..856J,1995MNRAS.274L..15P}.  

Spectral modeling (\S\ref{sec:spec-simult}) shows that the mass accretion rate of the system is very low.
In addition, the combined spectral and timing studies reveal a thermal state that dominated the accretion disc during the outburst of this system.

\section{Conclusion}
In this work, we performed a spectro-temporal analysis of the entire 2020 outburst of 4U 1755$-$338 using \textit{NICER} and \textit{NuSTAR} observations. Wideband spectral modeling is performed using the simultaneous \textit{NICER-NuSTAR} data to estimate the fundamental binary properties of the system. The main findings of the study are listed below.

\begin{itemize}
    \item The system 4U 1755$-$338 has a low interstellar neutral hydrogen column density, $N_H$= $0.443_{-0.007}^{+0.006}$ $\times$ 10$^{22}$ cm$^{-2}$.
    
    \item There is an evolving neutral medium surrounding the local environment of 4U 1755$-$338. The size of this partial covering absorber increases during the period of study and is identified as a growing cloud with constant density.

    \item The HID of the source does not follow the standard q-shaped pattern. The HID shows a correlated evolution of the hardness and the source flux, which indicates the thermal disc origin of the flux. Therefore, the HID of this source is very canonical to the soft state during the entire outburst, and the 2020 outburst of 4U 1755$-$338 can be considered as non-canonical.      

    \item Wideband spectral analysis performed in two modeling frameworks (based on \textit{kerrbb} and \textit{bhspec}) revealed the fundamental binary properties of the system. We identified from \textit{bhspec} (\textit{kerrbb}) based model that 4U 1755$-$338 is a low-luminosity system with an inclination of $67.44_{-3.03}^{+9.75}$ ($75.25_{-4.68}^{+5.59\circ}$) degrees, located at a distance of 8.5 \textit{kpc} that harbors a moderately spinning black hole with a spin parameter of $0.78_{-0.14}^{+0.02}$ ($0.50_{-0.43}^{+0.19}$) and  mass of $3.37_{-1.04}^{+0.45}\ (3.28_{-1.1}^{+1.7})M_{\odot}$ that accretes at a low ($0.28 \times 10^{18}$ $g/s$) rate.

    \item There is no evidence of reflection features in the spectra of 4U 1755$-$338.
       
    \item High $\Gamma$, $L \propto T^4$, behavior of HID, absence of reflection signatures, and weak variability in the PDS indicate that the source remains in the HSS during the outburst. 
\end{itemize}

\begin{acknowledgments}
\section*{Acknowledgments}
The authors thank the anonymous reviewer for their insightful comments and constructive suggestions, that significantly improved the quality of this publication. This work made use of data obtained from the \textit{NICER} and \textit{NuSTAR} missions operated by the \textit{National Aeronautics and Space Administration (NASA)}. We also acknowledge the use of software and services provided by the High Energy Astrophysics Science Archive Research Center (\texttt{HEASARC}) and \textit{NASA}’s Astrophysics Data System (\textit{ADS}) Bibliographic Services.
\end{acknowledgments}

\vspace{5mm}


\bibliography{reference}{}
\bibliographystyle{aasjournal}

\end{document}